\documentclass{PoS}

\title{Progress in developing a spiral fiber tracker for\\the J-PARC E36 experiment}

\ShortTitle{Progress in developing a spiral fiber tracker for the J-PARC E36 experiment}

\author{\speaker{Makoto Tabata}$^{,a,b}$, Keito Horie$^c$, Youichi Igarashi$^d$, Jun Imazato$^d$, Hiroshi Ito$^b$, Alexander Ivashkin$^e$, Hideyuki Kawai$^b$, Yury Kudenko$^e$, Oleg Mineev$^e$, Suguru Shimizu$^c$, Akihisa Toyoda$^d$, and Hirohito Yamazaki$^f$\\
\llap{$^a$}Institute of Space and Astronautical Science (ISAS), Japan Aerospace Exploration Agency (JAXA), Sagamihara, Japan\\
\llap{$^b$}Department of Physics, Chiba University, Chiba, Japan\\
\llap{$^c$}Department of Physics, Osaka University, Toyonaka, Japan\\
\llap{$^d$}Institute of Particle and Nuclear Studies (IPNS), High Energy Accelerator Research Organization (KEK), Tsukuba, Japan\\
\llap{$^e$}Institute for Nuclear Research (INR) of Russian Academy of Sciences (RAS), Moscow, Russia\\
\llap{$^f$}Research Center for Electron Photon Science, Tohoku University, Sendai, Japan\\
        E-mail: \email{makoto@hepburn.s.chiba-u.ac.jp}}

\abstract{This paper reports the recent progress made in developing a spiral fiber tracker (SFT) for use in the E36 experiment scheduled at the Japan Proton Accelerator Research Complex (J-PARC). The primary goal of this positive kaon decay experiment, which uses a stopped kaon beam, is to test lepton flavor universality to search for physics beyond the Standard Model of particle physics. For this experiment, we are currently upgrading the E246 apparatus, which consists of the superconducting toroidal spectrometer previously used at the High Energy Accelerator Research Organization (KEK), Japan. Conducting high-precision measurements will rely on efficiently detecting and tracking charged particles (i.e., positive muons and positrons) from kaon decays. Combined with the three existing layers of multiwire proportional chambers, the SFT comprises four layers of ribbons, with each layer containing 1-mm-diameter double-clad plastic scintillating fibers; the ribbons are spirally wound in two helicities around the kaon stopping target at the center of the detector system. Scintillation photons are read out by multipixel photon counters connected to the scintillating fibers by clear optical fiber extensions. A preliminary bench test shows that a prototype two-layer fiber ribbon exhibits 99.6\% detection efficiency at the 1-photoelectron threshold. Finally, the SFT was successfully assembled around the target holder.}

\FullConference{Technology and Instrumentation in Particle Physics 2014,\\
		2--6 June, 2014\\
		Amsterdam, the Netherlands}

\begin{document}

\section{Introduction}

We are currently developing a charged particle tracking device, which we call the spiral fiber tracker (SFT), for use in the E36 experiment \cite{cite1} scheduled at the Hadron Experimental Facility of J-PARC. This positive kaon decay experiment will search for physics beyond the Standard Model of particle physics by testing lepton flavor universality and searching for heavy sterile neutrinos and dark photons. To search for a violation of lepton flavor universality, we focus, in particular, on precisely measuring the ratio of the kaon decay widths $R_{\rm K} = \Gamma (K^+ \to ~e^+\nu)/\Gamma (K^+ \to ~\mu ^+\nu )$ using a stopped kaon beam. For this experiment, we are building a new TREK detector system by upgrading the detector \cite{cite2} used in the KEK-E246 experiment \cite{cite3}. To conduct high-precision measurements, it is vital to efficiently track charged particles (positive muons and positrons) from kaon decays and to identify the particles.

To reliably determine momentum requires at least four-point tracking, which suggests that segments must be tracked before and after the magnetic field generated by the detector system. Four independent tracking devices allow us to measure the detection efficiency of the charged decay particles by three-of-four-point tracking. The present E246 detector has three layers of multiwire proportional chambers. As a result, an additional tracking device is required. We therefore decided to develop and install the SFT in a limited space around a kaon stopping target at the center of the TREK detector.

\section{Design of spiral fiber tracker}

\begin{figure}[b] 
\centering 
\includegraphics[width=0.57\textwidth,keepaspectratio]{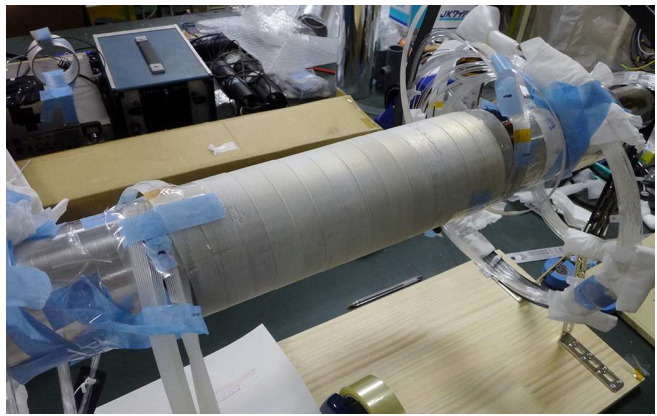}
\caption{Spiral fiber tracker before coiling the fourth layer.}
\label{fig:fig1}
\end{figure}

The heart of the SFT consists of ribbons, each containing a single layer of 1-mm-diameter double-clad plastic scintillating fibers (SCSF-78, Kuraray Co., Ltd., Japan). A total of four layers of ribbons, forming two sets of two layers in a staggered configuration, are spirally wound around the kaon stopping active target holder (approximately 20 cm; see Figure \ref{fig:fig1}). The two-layer combinations are coiled in two helicities and have different ribbon widths to phase shift the candidate positions of the charged particle hit. Fifteen (inner) or seventeen (outer) fibers, depending on the two helicities, are glued to form a single-layer flat ribbon approximately 5 m long. At both ends of each fiber, scintillation photons created in the fibers are read out by 128 multipixel photon counters (MPPCs). For each layer, approximately 6-m-long clear optical fibers are employed to extend the scintillating fibers to the upstream and downstream MPPCs with low transmission loss. To determine the real hit position from the candidate positions (i.e., ribbon turns), which are where the inner and outer fibers that detect scintillation photons cross each other, we shall use the supplementary information from hits or tracking on the active scintillating fiber target, in addition to timing information from both the ends of the SFT readout.

\section{Bench test of prototype fiber ribbon}

In March 2014, we bench-tested a prototype fiber ribbon glued by the Moderation-Line Co., Ltd., Japan to determine its charged particle detection efficiency. At that time, only a single ribbon was available. It was approximately 1.5 m long and consisted of sixteen 1-mm-diameter plastic scintillating fibers. At both the ends of the ribbon, where the fibers were not glued, the fiber ends were well polished, bundled, and connected to two photomultiplier tubes (PMTs; R9880U-210, Hamamatsu Photonics K.K., Japan). The MPPCs were not ready for this test. We used $\beta $ rays generated from a strontium-90 radio isotope as a source of minimum ionizing particles. The fiber ribbon was exposed to $\beta $ rays through a 5-mm-hole collimator fabricated from a 2-cm-long lead block. Event triggers were generated by coincidence signals from two scintillation counters positioned upstream and downstream of the ribbon. The upstream and downstream scintillation counters consisted of a multilayer sheet comprising 0.2-mm-diameter fibers and a 5-mm-thick block that a served as plastic scintillators, respectively.

\begin{figure}[b] 
\centering 
\includegraphics[width=0.57\textwidth,keepaspectratio]{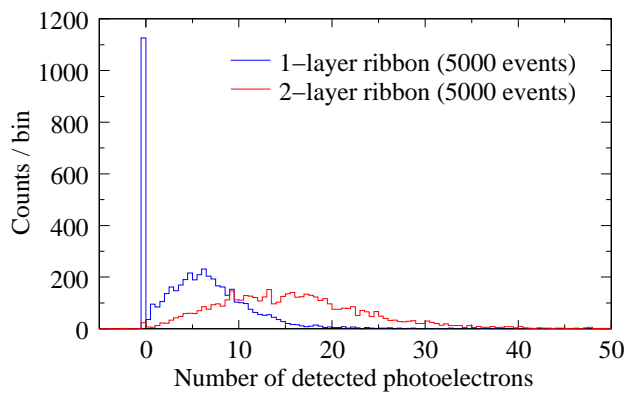}
\caption{Number distribution of photoelectrons detected by downstream PMT for (blue) single-layer and (red) two-layer ribbons.}
\label{fig:fig2}
\end{figure}

ADC spectra were obtained by computer-automated measurement and control (CAMAC) for single-layer and two-layer ribbons. To test the single-layer ribbon, we irradiated it with $\beta $ rays from a source positioned approximately 15 cm from one ribbon terminal. Here the PMT connected to the near end of the fiber and that connected to the opposite end of the fiber are called the ``upstream'' and ``downstream'' PMTs, respectively. The two-layer part of the ribbon was wound, so that the ribbon defined a cylinder. In the two-layer part, fibers were overlapped in a staggered configuration. The number distribution of photoelectrons detected by the downstream PMT is shown in Figure \ref{fig:fig2}. On average 15.2 photoelectrons were detected in the two-layer ribbon. For a threshold of 1 photoelectron at the upstream or downstream PMT, we obtained a detection efficiency of 99.6\% for the two-layer fiber ribbon and 78.9\% for the single-layer ribbon.

\section{Assembly of spiral fiber tracker}

In April 2014, we assembled the spiral fiber tracker at KEK. Plastic scintillating fibers were glued to form a total of four ribbons, and all fiber ends were connected to clear optical fibers by Moderation-Line Co., Ltd. To work on a desktop, we prepared a dummy target pipe with the same diameter (79 mm) as the actual target holder. First, the dummy pipe was wrapped by a thin Kapton film to facilitate removal of the coiled ribbons. Next the first- and second-layer ribbons were coiled in the staggered configuration around the Kapton sheet with left helicity. As shown in Figure \ref{fig:fig1}, the third- and fourth-layer ribbons were coiled with right helicity over the previous two layers. The coiled ribbons were fixed by Mylar tape. The ribbons and extended fibers were then shielded from light with a black sheet and black tubes, respectively. After that, to connect the fibers to the MPPCs, each clear fiber edge was glued to a coupler. Redundant glue was removed by polishing the fiber ends. Finally, the ribbons coiled around the Kapton sheet were transferred from the dummy pipe to the actual target holder.

\section{Conclusion}

We designed a new charged particle tracking device for use in the J-PARC E36 experiment. The device, called the spiral fiber tracker, is based on 1-mm-diameter plastic scintillating fibers and MPPCs and is designed to detect scintillation photons. A preliminary bench test of a prototype two-layer fiber ribbon indicates that the device would have a detection efficiency exceeding 99\% for the 1-photoelectron threshold. The actual spiral fiber tracker was assembled around the kaon stopping target holder.

\section*{Acknowledgments}

We are grateful to the members of the TREK/E36 collaboration for their assistance in this study. We are also grateful to Moderation-Line Co., Ltd. for their contributions to producing the fiber ribbon. This work was partly supported by the Russian Foundation Grant \#14-12-00560. M. Tabata was supported in part by the Space Plasma Laboratory at ISAS, JAXA.

\end{document}